\documentclass[prb,twocolumn]{revtex4-1}

\usepackage{amsmath}
\usepackage{amssymb}
\usepackage{amsthm}

\usepackage{graphicx}
\usepackage{graphics}
\usepackage{bm}
\usepackage{hyperref}
\usepackage{epsfig}
\usepackage{dcolumn}

\newcommand\vex[1]{\mathbf{#1}}

\def\sgn{\mathrm{sgn}}
\def\re{\mathrm{Re}\,}
\def\im{\mathrm{Im}\,}

\begin{document}

\title{Unpaired Majorana fermions in a layered topological superconductor}

\author{Babak Seradjeh and Eytan Grosfeld}

\address{Department of Physics, University of Illinois, 1110 West Green Street, Urbana, IL 61801-3080 USA}

\begin{abstract}
We study the conditions for the existence of unpaired Majorana modes at the ends of vortex lines or the side edges of a layered topological superconductor. We show that the problem is mapped to that of a general Majorana chain and extend Kitaev's condition for the existence of its nontrivial phase by providing an additional condition when a supercurrent flows in the chain. Unpaired Majorana bound states may exist in a vortex line that threads the layers if the spin-orbit coupling has certain in-layer components but, interestingly, only if a nonzero supercurrent is maintained along the vortex. We discuss the exchange statistics of vortices in the presence of unpaired Majorana modes and comment on their experimental detection.
\end{abstract}

\maketitle

%---- Introduction
\section{Introduction}

% Majorana fermions --
Majorana fermions were introduced in 1937~\cite{Maj37a} as neutral spin-$1/2$ particles that are their own anti-particles. Mathematically, they are described by a modified two-component Dirac equation, which admits purely real solutions. Majorana himself speculated that neutrinos might be described by this formalism. Indeed, after many years of hiatus, this appears now to be a promising suggestion~\cite{Wil09a}. Other candidate Majorana fermions are found in supersymmetric theories and theories of dark matter~\cite{Wil09a}. However, the experimental observation of Majorana fermions in these theories seems distant.

% condensed matter -- superconductivity
In recent years, an increasing number of proposals have been made for the realization of Majorana fermions as quasiparticles in condensed matter systems~\cite{ReaGre00a,Kit00a,FuKan08a,SauLutTew10a,LutSauDas10a}. While these proposals vary in their implementation, they generically exploit the special structure of quasiparticles in a superconductor, each of which is a mixture of an electron and a hole. By virtue of this structure, a quasiparticle whose energy is in the middle of the superconducting gap (a ``zero mode'' pinned to the chemical potential) is automatically a Majorana fermion. Hence, the aim of these proposals is to find systems where such zero modes can be stabilized. The existence of zero modes is a topological property of the Hamiltonian: they are robust against smooth changes in potentials. This topological protection would be an advantage in the experimental search for Majorana fermions once the right Hamiltonian is found or engineered.

\begin{figure}[t]
\begin{center}
\includegraphics[width=2.5in]{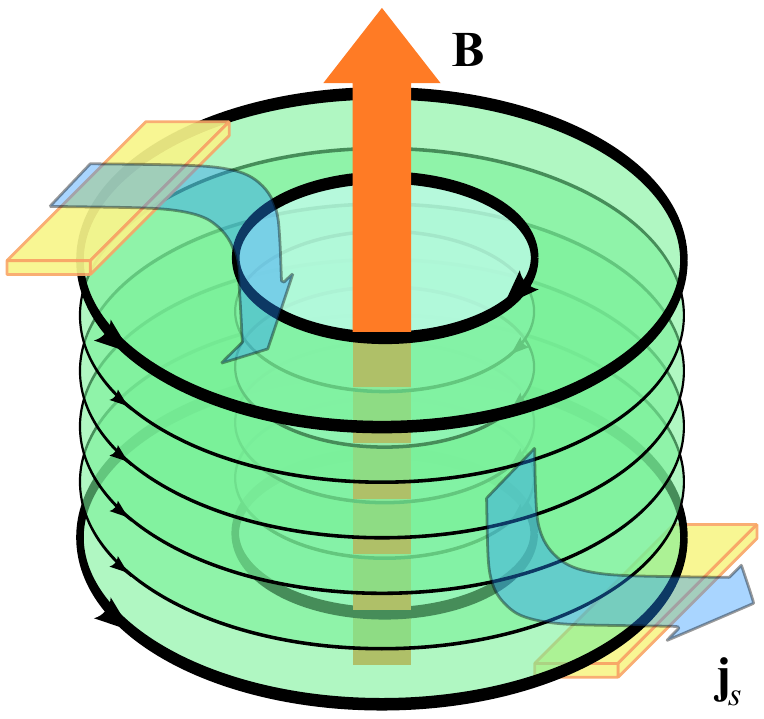}
\end{center}
\caption{(color online) The experimental setup with a multilayered topological superconducting annulus (green). The hole is threaded by the magnetic field $\vex B$ shown by the (orange) straight arrow. Superconducting leads, shown by (yellow) bars are attached to the top and bottom surface. A supercurrent $\vex j_s$ shown by (blue) curvy arrow, flowing along the hole, can be maintained between the leads. Majorana edge states are marked by arrows to show the direction of propagation. In the topological state, unpaired Majorana fermions are bound to edges at the top and bottom layer, shown by thick black circles.}
\label{fig:exp}
\end{figure}

% topological insulators -- SRO
A prime example is provided by a chiral p-wave ($p_x\pm  i  p_y$) superconductor in two dimensions~\cite{ReaGre00a}. A thin layer of the A-phase of superfluid $^3$He has this pairing symmetry. There is also evidence that Sr$_2$RuO$_4$, a layered material, may realize a similar pairing state in each layer~\cite{MacMae03a}. A close relative of this state, which combines both chiralities into a time-reversal invariant pairing state, also hosts Majorana modes protected by the additional time-reversal symmetry. The B-phase of superfluid $^3$He realizes this pairing symmetry. The doping-induced superconducting state of the topological insulator Cu$_x$Bi$_{2}$Se$_3$~\cite{Hor10a,Wra09a}, another layered material, might be another realization~\cite{FuBer10a}.  Vortices in these topological superconductors support Majorana bound states~\cite{KopSal91a}. In addition, the boundaries of these systems also contain propagating Majorana modes.

% quantum computation and layered problem --
In two spatial dimensions, the ground state of the superconductor with several Majorana bound states is highly degenerate and is unitarily mapped within the degenerate subspace upon exchange of two Majorana bound states. Therefore, these Majorana bound states behave as non-Abelian anyons~\cite{Iva01a} and could be useful for fault-tolerant quantum computation~\cite{Kit03a}. However, these properties may not survive in a layered system. Indeed, while two spatially separated Majorana fermions are protected, once they reach tunneling proximity they may hybridize into a regular fermion. So, if tunneling amplitude between Majorana fermions in and between layers is nonzero it might be expected that no unpaired Majorana fermions would survive.

% motivation etc. --
In this paper, we study the conditions under which unpaired Majorana fermions exist in  layered topological superconductors. Specifically, we study the Majorana fermions in a vortex line or the boundaries of the system normal to the layers (see Fig.~\ref{fig:exp}) creating a ``Majorana chain'' along the layers. In a spin-triplet chiral p-wave superconductor, a stable half-quantum vortex (HQV) carrying a flux $h/4e$ is allowed which supports a single Majorana bound state in each layer. Recently, strong evidence for the existence of HQVs in Sr$_2$RuO$_4$ has been reported~\cite{Raffi10}. A full-quantum vortex (FQV) carrying a flux $h/2e$, on the other hand, supports two Majorana bound states in each layer. The vortex in a time-reversal invariant topological superconductor also carries $h/2e$ flux and two Majorana bound states. The boundaries of both systems host two propagating Majorana modes per layer. Motivated by these observations and by a recent proposal to exploit $h/2e$ FQVs~\cite{GroSerVis10a}, we focus on the problem of two Majorana fermions per layer. For brevity we limit ourselves to the case of a chiral p-wave superconductor but our study may be readily generalized to other situations.

% advertise results --
Our main finding is that passing a supercurrent in the direction normal to the layers could stabilize unpaired Majorana bound states at the surface layers of the system. The equivalent Majorana chain  contains complex fermion tunneling amplitudes. Therefore, we generalize Kitaev's original work~\cite{Kit00a} to this case. Our result can be stated as follows. The intra-layer Majorana tunneling tends to pair the two Majorana fermions in each layer into a regular fermion and thus acts as a chemical potential. The interlayer tunneling amplitude, on the other hand, tends to pair Majorana fermions in different layers and acts as both complex tunneling as well as Cooper pairing of the regular fermions. The Majorana chain is then equivalent to a fermion chain in a superconducting state with a steady supercurrent. The condition for unpaired Majorana fermions at the chain boundaries is that the chemical potential lies in the fermion bandwidth (same as the condition derived by Kitaev) and that the supercurrent is below the critical current of the superconducting state of the fermion chain (in addition to Kitaev's condition).

In a layered chiral p-wave superconductor we find that, remarkably, unpaired Majorana fermions may only exist when a supercurrent is maintained normal to the layers. We shall discuss a set of sufficient conditions to stabilize such unpaired Majorana fermions. We explore their non-Abelian exchange statistics and also discuss ways to detect them. Details of our calculations are presented in two appendices.

%---- Topological phases of the layered system
\section{The Majorana chain}

Let us start with the following problem: consider a one-dimensional lattice with sites $\ell=1,2,\cdots N$ and two Majorana fermions $\gamma_{\sigma\ell}$, $\sigma\in\{\uparrow,\downarrow\}$ at each site. The most general Hamiltonian describing Majorana tunneling at each site and between nearest neighbors is
%%%%%%
\begin{equation}\label{eq:HtMajorana}
%H_{\rm t} = i\sum_{\ell=1}^{N-1}\left(t_{\ell\uparrow} \gamma_{\ell\uparrow}\gamma_{\ell+1\uparrow} + t_{\ell\downarrow} \gamma_{\ell\downarrow}\gamma_{\ell+1\downarrow} + t_{\ell\uparrow\downarrow} \gamma_{\ell\uparrow}\gamma_{\ell+1\downarrow} + t_{\downarrow\uparrow} \gamma_{\ell\downarrow}\gamma_{\ell+1\uparrow}\right) + i\sum_{\ell=1}^N \alpha \gamma_{\ell\uparrow}\gamma_{\ell\downarrow}.
H_{\rm t} =   i \sum_{\ell=1}^{N-1} \sum_{\sigma\sigma'} t_{\sigma\sigma'\ell} \gamma_{\sigma\ell}\gamma_{\sigma'\ell+1} +  i \sum_{\ell=1}^N \alpha_\ell \gamma_{\uparrow\ell}\gamma_{\downarrow\ell},
\end{equation}
%%%%%%
with $t_{\sigma\sigma'\ell}, \alpha_\ell \in \mathbb{R}$. The Majorana operators satisfy the algebra $\{ \gamma_{\sigma\ell}, \gamma_{\sigma'\ell'} \} = 2 \delta_{\sigma\sigma'} \delta_{\ell\ell'}$. Let us define a fermionic operator for each layer, $\psi_\ell = \frac12(\gamma_{\uparrow\ell} +  i  \gamma_{\downarrow\ell})$ which satisfy the usual fermionic algebra $\{ \psi_\ell^\dagger , \psi_{\ell'} \} = \delta_{\ell\ell'}$ and $\{ \psi_\ell,\psi_{\ell'}\} = 0$. Then
%%%%%%
\begin{equation}\label{eq:HtFermi}
H_{\rm t} = \sum_{\ell=1}^{N-1} \left( t_\ell \psi^\dagger_{\ell}\psi_{\ell+1} + \Delta_\ell \psi_{\ell}\psi_{\ell+1} + \mathrm{H.c.} \right) - \sum_{\ell=1}^N \mu_\ell \psi^\dagger_\ell\psi_\ell + c,
\end{equation}
%%%%%%
where
%%%%%%
\begin{subequations}
\begin{eqnarray}
t_\ell &=& t_{\uparrow\downarrow\ell}-t_{\downarrow\uparrow\ell} +  i (t_{\uparrow\uparrow\ell}+t_{\downarrow\downarrow\ell}) \equiv |t_\ell|  e^{ i \theta_\ell}, \\
\Delta_\ell &=& t_{\uparrow\downarrow\ell}+t_{\downarrow\uparrow\ell} +  i (t_{\uparrow\uparrow\ell}-t_{\downarrow\downarrow\ell}) \equiv |\Delta_\ell|  e^{ i \phi_\ell}, \\
\mu_\ell &=& -2\alpha_\ell,
\end{eqnarray}
\end{subequations}
%%%%%%
and $c = -\sum_{\ell=1}^N \alpha_\ell$ is a constant.

\begin{figure}[tb]
\begin{center}
\includegraphics[width=3.2in]{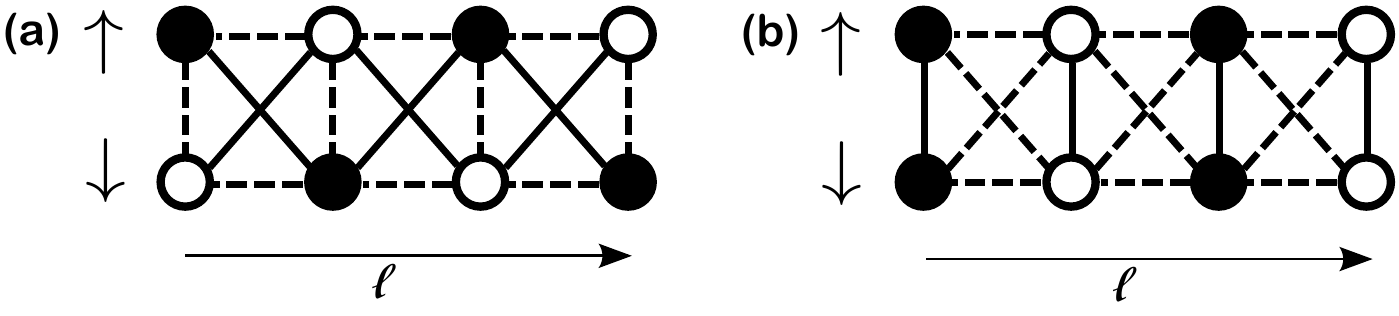}
\end{center}
\caption{The sign assignments $\gamma_{\sigma\ell}\to\pm\gamma_{\sigma\ell}$, $\sigma\in\{\uparrow,\downarrow\}$ of Majorana operators for $\mathbb{Z}_2$ spectral symmetries. Black and white circles indicate opposite sign assignments. Dashed (solid) lines show tunneling amplitudes that are multiplied with a $-$ ($+$) sign, (a) $\mu_\ell\to-\mu_\ell, t_\ell\to t_\ell^*$, $\Delta_\ell\to\Delta_\ell^*$; and (b) $t_\ell\to-t_\ell$, $\Delta_\ell\to-\Delta_\ell$.}
\label{fig:pmMajorana}
\end{figure}

The spectral properties of Hamiltonians~(\ref{eq:HtMajorana}) and~(\ref{eq:HtFermi}) are invariant under the following operations: (a) $\mu_\ell\to-\mu_\ell, t_\ell\to t_\ell^*$, $\Delta_\ell\to\Delta_\ell^*$; and (b) $t_\ell\to-t_\ell$, $\Delta_\ell\to-\Delta_\ell$. In each case, the changes can be undone by a redefinition of the Majorana operators $\gamma_{\sigma\ell}\to\pm\gamma_{\sigma\ell}$. The choice of signs is illustrated in Fig.~\ref{fig:pmMajorana}. %These symmetries will be used in the following. 

In general, we can remove the complex phase of $t_\ell$ by a gauge transformation
%%%%%%
\begin{subequations}
\begin{eqnarray}
\psi_\ell\to  e^{ i \eta_\ell}\psi_\ell, &\quad& \eta_{\ell}-\eta_{\ell+1}=\theta_\ell, \\
\Delta_\ell \to \Delta_\ell  e^{ i (\eta_\ell+\eta_{\ell+1})} &\quad& \therefore \phi_\ell \to \phi_\ell + (\eta_\ell+\eta_{\ell+1}).
\end{eqnarray}
\end{subequations}
%%%%%%
It follows from the last equation that there is a $\theta_\ell+\theta_{\ell+1} ({\rm mod}\ 2\pi)$ contribution to the interlayer supercurrent, $j_{{\rm s},\ell} \propto \phi_{\ell+1}-\phi_\ell$, which is nonzero when $t_\ell\not\in\mathbb{R}$. 

From here on let us assume that all amplitudes are constant, $t_\ell=t, \Delta_\ell=\Delta, \mu_\ell=\mu$. Then, the spectrum of~(\ref{eq:HtFermi}) as a function of the lattice momentum $q$ is
%%%%%%
\begin{equation}
E_{q,\pm} = (\im t) \sin q \pm \sqrt{\left[(\re t) \cos q - \mu/2\right]^2 + |\Delta|^2\sin^2 q},
\end{equation}
%%%%%%
which has the symmetry $E_{-q,-} = - E_{q, +}$. We can see from this expression that the $\im(t)$ shifts the energies and therefore gives rise to a current.

The Hamiltonian in~(\ref{eq:HtFermi}) when $t\in\mathbb{R}$ was considered by Kitaev~\cite{Kit00a}, who showed that it admits two phases. In the trivial phase, there are no unpaired Majorana modes. In the nontrivial phase, there are unpaired Majorana modes at the boundaries of the system. The nontrivial phase exists when~\cite{Kit00a}
%%%%%%
\begin{equation}\label{eq:endM}
|\mu| < 2|\re t|,
\end{equation}
%%%%%%
which is the condition for a partially filled band.

We now show that when $t\not\in\mathbb{R}$, the nontrivial phase obtains when in addition to~(\ref{eq:endM}),
%%%%%%
\begin{equation}\label{eq:super}
|\Delta| > |\im t|.
\end{equation}
%%%%%%
The quantity $|\Delta|-|\im t|$ is proportional to the gap at the Fermi points where $(\re t)\cos q_F=\mu/2$. Physically, this condition ensures that the supercurrent is less than the critical value above which superconducting gap in~(\ref{eq:HtFermi}) is destroyed: the complex phase of $t$ is restricted to $|\tan\theta| < 2|\Delta|/|\mu|$, which guarantees that the supercurrent $j_s \propto 2\theta$ (for small $|\Delta|$) is less than the critical current in the chain~\cite{Tinkham}. This is one of our main results.

To see this, we pass to the continuum limit of~(\ref{eq:HtMajorana}), by writing $z=\ell a$, $a\to0$. There are two counter-propagating low-energy modes with lattice momenta $q\approx0$ and $q\approx\pi/a$ where the gap vanishes. Expanding around their respective momenta we have $\gamma_{\sigma\ell+1}\approx \pm[\gamma_{\sigma}(z) + \partial_z\gamma_{\sigma}(z)]$. So,
%%%%%%
\begin{subequations}
\begin{eqnarray}
H_{\rm c} &=& \frac{ i }2 \int  d  z \left[\xi_0^T \left( \Gamma \partial_z + M  \right) \xi_0  + \xi_\pi^T \left( - \Gamma \partial_z + M_\pi \right) \xi_\pi \right], \nonumber\\
~\\
\Gamma &=& (\im t) I + (\im \Delta) \sigma_z +(\re \Delta) \sigma_x, \\
M &=& \left(\re t - \frac12\mu\right)  i \sigma_y, \quad M_\pi = \left(- \re t - \frac12\mu\right)  i \sigma_y,
\end{eqnarray}
\end{subequations}
%%%%%%
with $\xi_{0 (\pi)}=(\gamma_\uparrow, \gamma_\downarrow)^T$ for modes near $0$ ($\pi$). We now look for zero-energy solutions of the Dirac operator $D=\Gamma\partial_z + M$ localized where the mass $m \equiv \re t - \frac12\mu$ changes sign. The localized mode corresponds to an unpaired Majorana bound state at the boundary of the nontrivial phase. Let us take $m=|m|\sgn(z)$ for simplicity;  then, if
%%%%%%
\begin{equation}
\epsilon\equiv  |\Delta|^2 - \left(\im t\right)^2 >0,
\end{equation}
%%%%%%
we find $D e^{-k|z|}\xi = 0$ with $k = |m|/\sqrt{\epsilon}$, and
%%%%%% 
\begin{equation}
\xi = 
\left( \begin{array}{c} \sgn(m)\sqrt{\epsilon} - \re\Delta  \\ \im(\Delta+t) \end{array} \right) ~ {\rm or} ~ 
\left( \begin{array}{c} \im(\Delta-t) \\ \sgn(m)\sqrt{\epsilon} + \re\Delta \end{array} \right), 
\end{equation}
%%%%%%
whichever is nonzero.
For modes near $\pi$ the Dirac operator is $D_\pi = -\Gamma\partial_z + M_\pi$. So, for $\mu\neq0$ the mass changes sign either for the modes near $0$ or the ones near $\pi/a$ and there is only one zero mode. Since the sign of $t$ and $\mu$ can be changed by appropriate sign assignments of the Majorana operators, we find that~(\ref{eq:endM}) gives the condition for the nontrivial phase also when $t\not\in\mathbb{R}$ subject to the additional condition~(\ref{eq:super}).

\begin{figure}[tb]
\begin{center}
\includegraphics[width=3.1in]{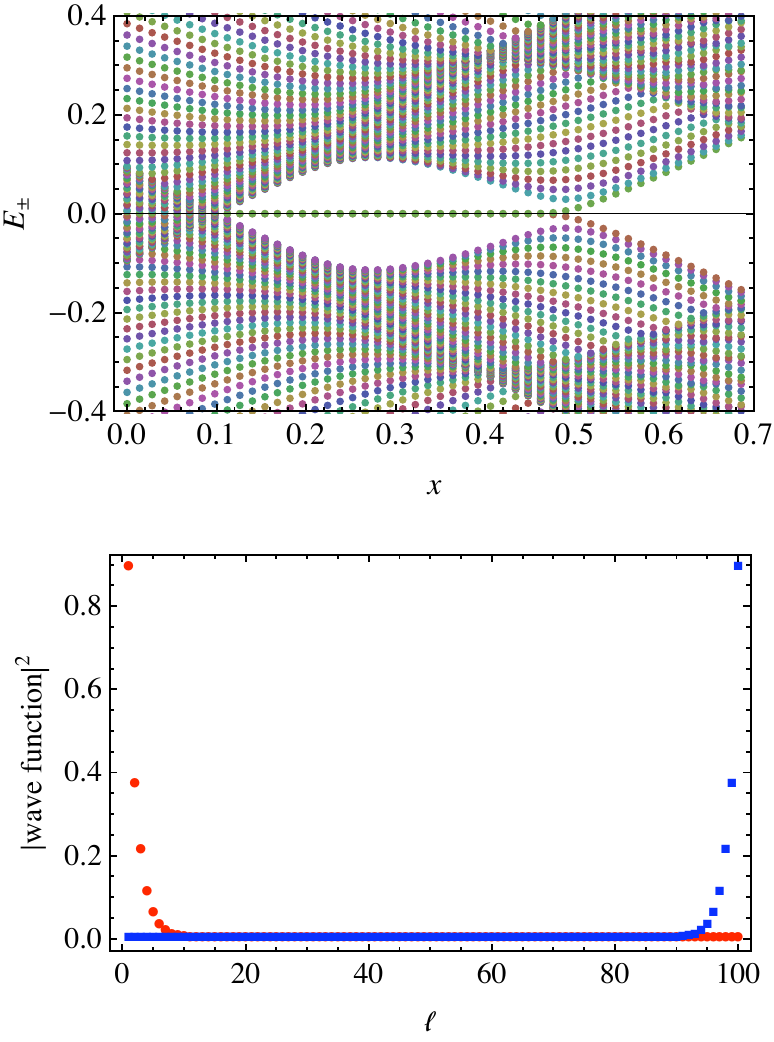}
\end{center}
\caption{(color online) The spectrum and eigenstates for a Majorana chain with $N=100$ sites. 
Top: The spectrum of the Majorana chain as function of $x\equiv t_{\uparrow\downarrow}$ (not all eigenvalues are shown). The other parameters are fixed as $
\alpha=0.2$, $t_{\uparrow\uparrow}=t_{\downarrow\downarrow}=0.4$, $t_{\downarrow\uparrow}=0.7$. The conditions (\ref{eq:endM}) and (\ref{eq:super}) yield $x<0.5$ and $x>0.1$, respectively. Two zero modes are clearly visible in the spectrum for $0.1<x<0.5$. 
Bottom: The wave functions of the two zero modes for $x=0.25$. Squares (blue) and circles (red) distinguish the two modes. The Majorana modes are localized at the two ends of the chain.}
\label{fig:numerics}
\end{figure}

We have also confirmed these results numerically. In Fig.~\ref{fig:numerics}, we present the results of exact diagonalization of Hamiltonian (\ref{eq:HtMajorana}) on a chain with $N=100$ sites. The parameters are chosen to explore the conditions (\ref{eq:endM}) and (\ref{eq:super}) for the presence of Majorana modes. It is clear that both conditions are necessary (and sufficient) for the existence of unpaired Majorana modes.

%To see how these considerations might be physically relevant we now turn to the layered chiral p-wave superconductor.

We now turn to study in detail the case of a chiral p-wave superconductor of the type that is thought to be realized in Sr$_2$RuO$_4$. In this state, while the time-reversal symmetry is spontaneously broken by the order parameter, the presence of spin degeneracy gives rise to two Majorana modes per layer. These Majorana modes may then couple both within a layer and in between layers through the Zeeman term, the spin-orbit term, and direct wave function overlaps between layers. We show that in a finite range of parameters a topological state is established in the vortex line (see Fig.~\ref{fig:phased}). %The second case is a time-reversal invariant system of the type that may be realized in a stack of thin layers of $^3$He-B, where in the presence of a vortex two Majorana modes appear per layer. This pair of Kramers' degenerate states will not couple easily, and the few allowed tunneling terms will always lead to a topologically trivial state, see Fig. .

%---- Intralayer and interlayer Majorana tunneling
\section{Unpaired Majorana bound states in a vortex line}

The quasiparticles of the superconducting state are described by the Bogolubov--de~Gennes (BdG)  Hamiltonian $\Psi^\dag H \Psi/2$, where the Nambu operator $\Psi=(\psi,\psi^\dag)^T$, $\psi$ is the electron annihilation operator,
\begin{eqnarray}
	H=\left(\begin{array}{cc} h & \Delta \\ \Delta^\dag & -h^T\end{array}\right),
\end{eqnarray}
with $h$ the Hamiltonian density of the system in the absence of superconductivity, and $\Delta$ the pairing order parameter of the condensate. We have suppressed the spin index for brevity.  The order parameter is skew-symmetric $\Delta^T=-\Delta$, therefore $\tau_x H^* \tau_x = - H$ where $\tau_x$ is the Pauli matrix acting in the Nambu space. The BdG quasiparticle of energy $E$ is annihilated by $\Psi_E = \sum \chi_{E}^T \Psi$, where $H\chi_E=E\chi_E$. By the symmetry of the BdG Hamiltonian, $\chi_{-E}=\tau_x\chi_{E}^{*}$. Noting that by construction $\Psi^\dagger=\Psi^T\tau_x$, we have $\Psi_E^\dagger=\Psi_{-E}$. In particular, at zero energy, $\Psi_{0}=\Psi_{0}^{\dag}\equiv\gamma$ would be a Majorana fermion. 

For a spin-triplet chiral p-wave superconductor the pairing term has the form
\begin{eqnarray}\label{eq:triplet} 
	\Delta=\frac{ i }{2}v_\Delta  e^{ i \Phi/2}\{\partial_x- i \partial_y,({\boldsymbol \sigma}\cdot {\mathbf d})\sigma_y\} e^{ i \Phi/2},
\end{eqnarray}
%%%%%%
where $\Phi$ is the phase of the order parameter, and $v_\Delta$ is a constant gap velocity.
Here $\vex d$ determines the direction of pairing in the space spanned by the spin-triplet states, and $\{\bullet,\bullet\}$ is the anticommutator. Restoring the spin indices, now $\psi=(\psi_\uparrow, \psi_\downarrow)^T$. 

%The Hamiltonian is invariant under a rotation by angle $\phi$ around the z-axis, compensated by $\Phi\to\Phi+\phi$, and a suitable rotation of $\vex d$. This allows us to choose a constant $\vex d$ to be along the y-axis.

The Hamiltonian is symmetric  under the mapping ${\mathbf d}\to -{\mathbf d}$ and $\Phi\to \Phi+\pi$. This $\mathbb{Z}_2$ symmetry stabilizes a half-quantum vortex. When encircling a HQV, the phase of the order parameter changes by $\pi$, so $\Phi=\phi+\theta/2$ where $\phi$ is the smooth part of the phase. The extra angular momentum needed to produce a single-valued wave function is provided by rotating $\vex d$ to $-\vex d$. For simplicity, we consider a configuration where $\vex d$ rotates in the xy-plane, say $\vex d= (\cos\theta/2,-\sin\theta/2,0)$. For this configuration, there is a $2\pi$ winding of the oder parameter phase only in the spin-$\uparrow$ block of the BdG Hamiltonian. Therefore, there is a single zero mode with support in spin-$\uparrow$ component in each layer and an unpaired Majorana bound state, $\gamma_\uparrow$ (see Appendix~\ref{app:Mmodes} for details).

In the full quantum vortex $\vex d$ is constant and the phase of the order parameter winds by $2\pi$. Both spin components now see the phase winding and there are, therefore, two zero modes per layer corresponding to two Majorana bound states, $\gamma_\uparrow$ and $\gamma_\downarrow$.

The Majorana fermions can tunnel between the layers and, in the case of the FQV, within each layer. Here we consider first-order perturbations that break spin rotation symmetry, namely the Zeeman splitting, $h_{\rm Z}=\mu_{\rm B}\,\boldsymbol{\sigma} \cdot {\vex B}$, with an effective Bohr magneton $\mu_{\rm B}$, and the spin-orbit interaction $h_{\rm so}=\boldsymbol\lambda \cdot (\boldsymbol{\sigma}\times \vex p)$ with the spin-orbit coupling $\boldsymbol\lambda $. A summary of the results follows with details given in Appendix~\ref{app:tun}. %We have used the symmetries of the Hamiltonian to choose $\vex d$ along the y-axis.

The intra-layer tunneling due to the spin-orbit interaction is zero since the zero energy solutions in the same layer have equal orbital angular momentum. The intra-layer Zeeman tunneling also vanishes when $\vex B \perp \vex d$. This is true to all orders of perturbation theory, since in this case, the effect of the Zeeman term is to shift the chemical potentials of each spin component by $\pm \mu_{\rm B} B$. Since the two spin blocks are decoupled, we still find two zero modes.

%The interlayer tunneling amplitude is given by the overlap of wave functions in different layers. This is exponentially suppressed by a factor $ e^{-a/a_0}$, where $a$ is the inter-layer separation and $a_0$ is a characteristic length scale. 

\begin{figure}[t]
\begin{center}
\includegraphics[width=1.65in]{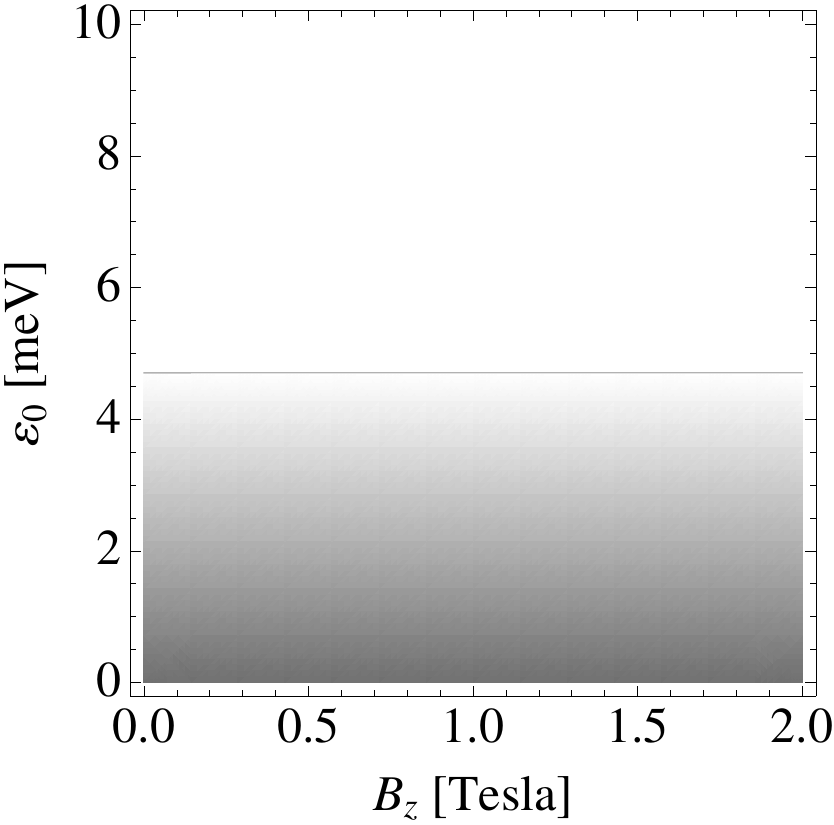}
\includegraphics[width=1.65in]{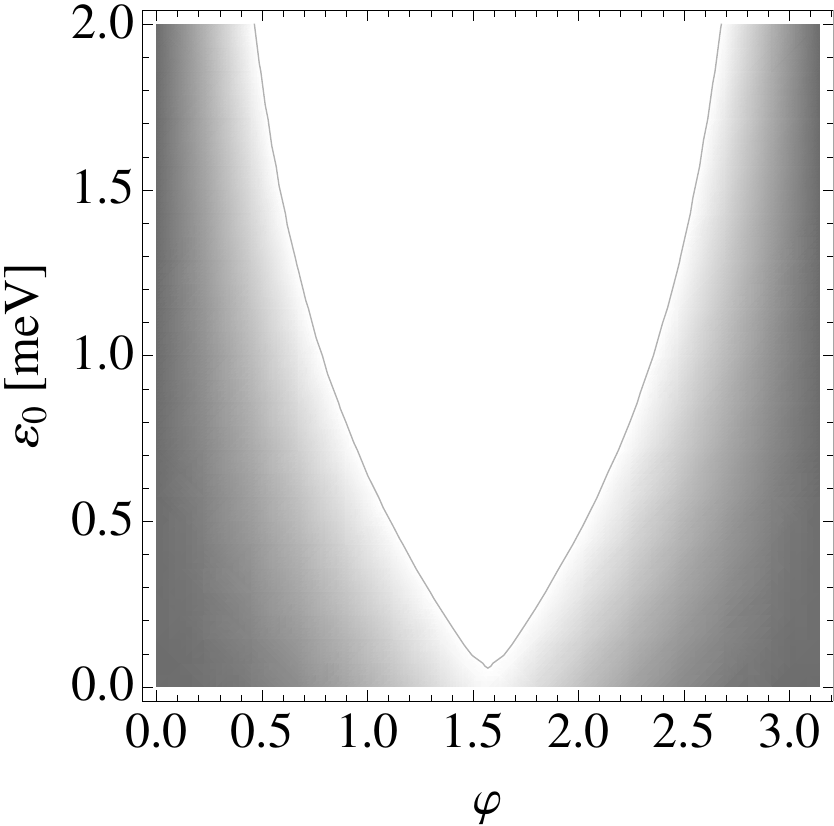}

\vspace{2mm}
\includegraphics[width=1.65in]{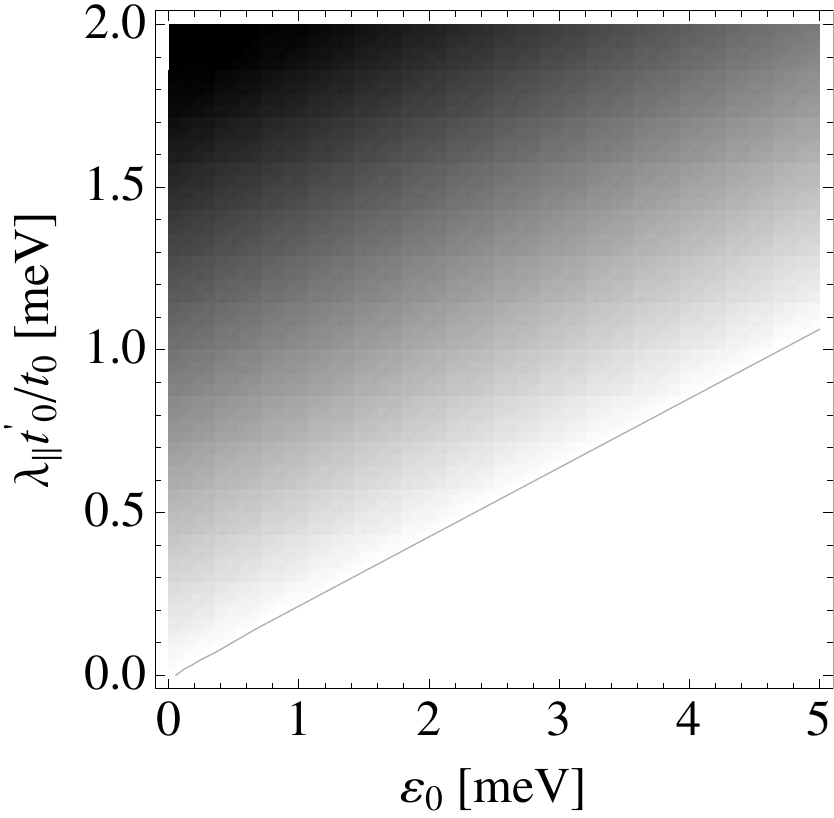}
\includegraphics[width=1.65in]{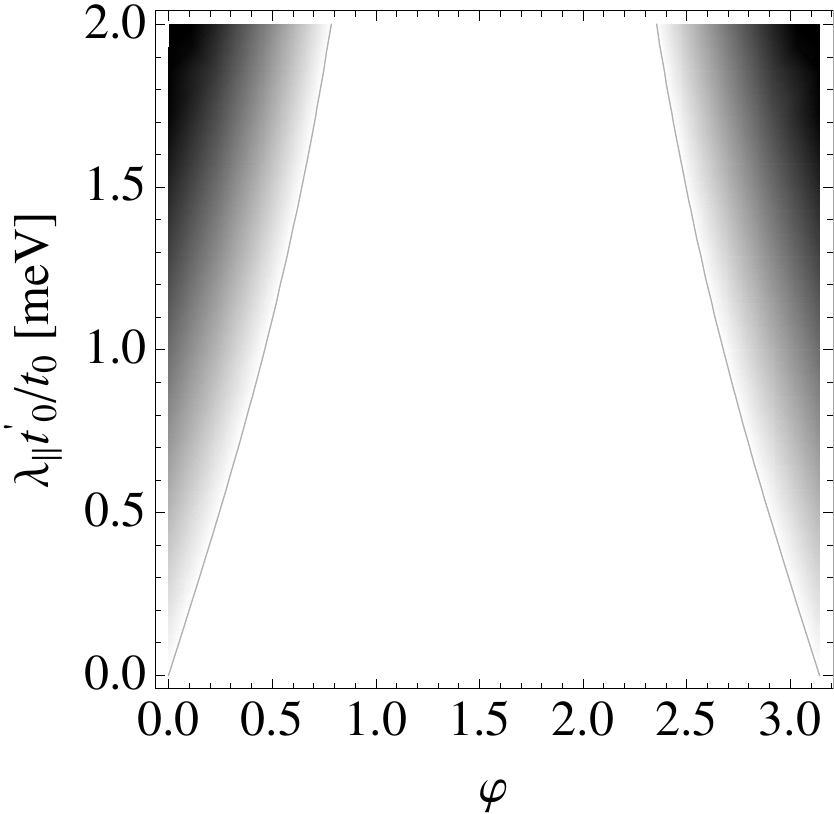}

%\vspace{2mm}
%\includegraphics[width=1.65in]{fig4e.pdf}
%\includegraphics[width=1.65in]{fig4f.pdf}

%\begin{picture}(1,1)
%\put(77,88){\myblack{${\bf j}_s$}}
%\put(7,146){\myred{${\bf J}_v$}}
%\put(-3,91){\myblue{$Q$}}
%\put(-53,82){\myred{$L$}}
%\put(47,82){\myred{$R$}}
%\end{picture}
\end{center}
\caption{The phase diagram for $\vex d=\hat{\vex d}_\perp, \vex B=B_z\hat{\vex z}$. The shaded regions are topologically nontrivial which support unpaired Majorana bound states. In each panel the two parameters that are not shown take values as $\varepsilon_0=2$~meV, $B_z=1$~Tesla, $\lambda_\parallel\equiv\boldsymbol\lambda\cdot\vex d=(t_0/t'_0)\times 1$~meV, and $\phi=\pi/15$. Over the a range of a few Tesla the Zeeman energy is negligible, as seen in the top left panel. (We have taken the effective Bohr magneton to be the same as the bare value $5.7\times 10^{-2}$~meV$/$Tesla.) The gray scale shows the value of $|\Delta|-|\im t|\sim$ the energy gap, where white is zero and black is $t_0\times 3.5$~meV. The line indicates where the gap vanishes.}
\label{fig:phased}
\end{figure}

The tunneling amplitudes are
%%%%%%
\begin{subequations}
\begin{eqnarray}
\re t &=& - 2 t_0 \mu_{\rm B} \vex B\cdot\vex d \cos \varphi + 2 t'_0 (\vex d \times \boldsymbol\lambda )_z \sin\varphi, \\
\im t &=& 2t_0 \varepsilon_0 \sin \varphi, \\
\re \Delta &=& 2t'_0 \boldsymbol\lambda \cdot\hat{\vex d}_\perp \cos\varphi + 2t_0 \mu_{\rm B} (\vex B \times \hat{\vex d}_\perp)_z \sin\varphi,  \\
\im \Delta &=& 2t'_0  \hat{\vex d}_\perp \cdot(\vex d \times \boldsymbol\lambda ) \cos\varphi  \nonumber\\
&& +\  2t_0 \mu_{\rm B} \left[\hat{\vex d}_\perp \times (\vex B \times \vex d)\right]_z \sin\varphi,\\
\mu &=& 4 \mu_{\rm B} \vex B\cdot \vex d.
\end{eqnarray}
\end{subequations}
%%%%%%
Here $\varphi = (\Phi_1 - \Phi_2)/2$ is half of the superconducting phase difference between the layers, $\varepsilon_0$ is the energy scale of direct electron tunneling between the layers, and $t_0$ and $t'_0$ are, respectively, direct and differential overlap integral between the layers. We use the notation that $\overline{\sigma}$ is the opposite of $\sigma$, the signs $(-1)^\uparrow=1=-(-1)^\downarrow$, and $\hat{\vex d}_\perp$ is the unit vector along the projection of $\vex d$ onto the xy-plane. 

When $\varphi=0$, we see that $\re t = - 2 t_0 \mu_{\rm B} \vex B\cdot\vex d$, $\im t =0$, $\re\Delta = 2 t'_0 \boldsymbol\lambda \cdot\hat{\vex d}_\perp$ and $\im \Delta=2 t'_0  \hat{\vex d}_\perp \cdot(\vex d \times \boldsymbol\lambda )$. There is no supercurrent, but Kitaev's condition~(\ref{eq:endM}) yields $\vex B\cdot \vex d\neq0$ and $|t_0| > 1$, which is unlikely to be satisfied since  $t_0 \propto  e^{-a/a_0}$.

For $\varphi\neq 0$, it becomes possible to satisfy Kitaev's condition. The supercurrent is also nonzero in this case. Kitaev's condition holds if
%%%%%%
\begin{equation}\label{eq:B.d}
\mu_{\rm B}|\vex B \cdot \vex d| \ll t'_0\left|(\vex d \times \boldsymbol\lambda )_z\sin\varphi\right|,
\end{equation}
%%%%%%
which dictates that $\vex d$ must have nonzero components in the xy-plane and that $\boldsymbol{\lambda}$ also have components normal to $\hat{\vex d}_\perp$ in the xy-plane. The additional condition~(\ref{eq:super}) for the supercurrent could also be satisfied if either (say, for small $\varphi$)
%%%%%%
\begin{equation}\label{eq:lambdaso}
|t'_0\boldsymbol\lambda \cdot\hat{\vex d}_\perp| > |t_0\varepsilon_0| \quad{\rm and/or}\quad
|t'_0\hat{\vex d}_\perp \cdot(\vex d \times \boldsymbol\lambda )| > |t_0\varepsilon_0|,
\end{equation}
%%%%%%
which means that $\boldsymbol{\lambda}$ should have some components in the xy-plane, or
%%%%%%
\begin{equation}\label{eq:Bxd}
\mu_{\rm B} |\vex B \times \hat{\vex d}_\perp| > |\varepsilon_0|.
\end{equation}
%%%%%%
Since Zeeman energy is generally very small ($\sim 10^{-5}$~eV at 1 Tesla) the last condition would require fine tuning to hold.

Our discussion suggests that unpaired Majorana modes may be stabilized if a supercurrent flows between the layers, $\vex d$ is unlocked from the z-axis, $\vex B\cdot \vex d \to 0$, and there is a spin-orbit coupling in the xy-plane. To illustrate let us take $\vex d=\hat{\vex d}_\perp$ to be in the xy-plane, and $\vex B = B_z \hat{\vex z}$. Then (\ref{eq:B.d}) is automatically satisfied if $\boldsymbol\lambda $ has normal components to $\vex d$ in the xy-plane. If in addition $\boldsymbol\lambda $ has components along $\vex d$, it  is possible to satisfy (\ref{eq:lambdaso}) for large enough spin-orbit coupling. See Fig.~\ref{fig:phased}.

%---- Effects on exchange statistics
\section{Exchange statistics of vortices}

Majorana bound states in superconducting vortices act as non-Abelian anyons in a single layer~\cite{Iva01a}. The reason is that when a vortex is taken around another in a loop, the superconducting phase changes by $2\pi$. Therefore the Majorana bound states $\gamma', \gamma$ at the vortices each acquire a minus sign.  This is generated by the transformation  $U'=\gamma'\gamma $ on the zero-energy Hilbert space. Let us now consider an experiment where a beam of vortices with Majorana bound states $\{ \dots \gamma'',\gamma' \}$ (say, HQVs in a p-wave superconductor) circumvents a stationary vortex in the middle with Majorana bound state $\gamma$. Using the anticommutation of Majorana operators, we can see that for two incoming vortices $U''$ and $U'$ do not commute. In fact, they anticommute. In a vortex interference experiment, this non-Abelian structure makes the interference term vanish~\cite{SteHal06a,OveBai01a}.

What happens in the layered case? The vortex is now an extended object with many Majorana bound states. When it encircles another vortex, the Majorana fermions on each layer acquire a minus sign, generated by
%%%%%
\begin{equation}
	\label{eq:U-HQV}
	U'_N=\prod_{\ell=1}^N \gamma'_\ell \gamma_\ell.
\end{equation}
%%%%%
It is easy to see that $U''_N$ and $U'_N$ do not commute only when $N$ is odd. This ``odd-even'' effect in the number of layers is a robust feature that depends on the existence or not of an unpaired Majorana fermion within the loop.

Now let us consider a moving FQV with two Majorana bound states in each layer. First we  look at the case when all Majorana tunneling are turned off. When the FQV goes around a stationary HQV with a single Majorana bound state per layer in, say, $\gamma_{\uparrow\ell}$, it was recently shown~\cite{GroSerVis10a} that $\gamma_{\uparrow\ell}\to-\gamma_{\uparrow\ell}$ while $\gamma_{\downarrow\ell}\to\gamma_{\downarrow\ell}$. The generator of this transformation is then,
$ 
U'_{\uparrow N}=\prod_{\ell=1}^N\gamma'_{\uparrow\ell} \gamma_{\uparrow\ell}.
$
Since this is in the same form as~(\ref{eq:U-HQV}), the results are also the same: the exchange of a FQV and a HQV is non-Abelian only when $N$ is odd.

When Majorana tunneling amplitudes are turned on, the situation is quite different. Now, the Majorana chain in the FQV has two topologically distinct phases. In the trivial phase, all the Majorana fermions are paired into regular fermions that are all either occupied or unoccupied. Therefore, in this case the Majorana states become decoherent and non-Abelian statistics is lost over a long enough time. In the nontrivial phase, there are unpaired Majorana fermions at the ends of the vortex lines on the surface of the superconductor. Then, the non-Abelian statistics between a FQV and a HQV is recovered for odd $N$ similar to the case of a single layer. An even more interesting, but rather speculative situation is when the vortex is partially pinned so that only one of the end Majorana fermions can move, in which case non-Abelian statistics is obtained at a single surface layer regardless of the evenness or oddness of $N$ both between FQVs and between a FQV and a HQV.

%--- Experiment
\section{Experimental remarks}

%\begin{figure}[t]
%\begin{center}
%\includegraphics[width=2.5in]{fig5.pdf}
%\end{center}
%\caption{(color online) A possible experimental setup. The layered (green) topological superconductor is punched with hole (transparent) threaded by the magnetic field $\vex B$ shown by the (orange) straight arrow. Superconducting leads, shown by (yellow) long cubes are attached to the top and bottom surface. A supercurrent $\vex j_s$ shown by (blue) curvy arrow, flowing along the hole, can be maintained between the leads. In the topological state, unpaired Majorana fermions are bound to the hole boundaries at the top and bottom layer, shown by thick black circles.}
%\label{fig:exp}
%\end{figure}

Since the unpaired Majorana states discussed in this work appear at the surfaces of the sample, one might expect that surface sensitive probes such as scanning tunneling spectroscopy would be able to detect them. The tunneling spectrum would show an enhanced resonance at zero bias, which exponentially decays away from the Majorana bound state. A possible setup is shown in Fig.~\ref{fig:exp}. When the topological superconductor has an even number of layers, there is in total always an even number of Majorana fermions bound to the hole boundaries. In the trivial phase, all Majorana fermions are paired up and there will be no zero-bias resonance. In the nontrivial phase, however, two of these Majorana fermions are bound to the top and bottom layers and are separated by an energy gap from the rest of the spectrum. The tunneling spectrum would then show a zero-bias resonance. As we have discussed, such a state may be obtained in a chiral p-wave superconductor when a supercurrent flows between the leads and along the hole. 

A fine energy resolution is needed to detect the zero-bias Majorana resonance. First, one needs to be able to resolve the mini-gap $\sim v_\Delta/R$, separating the Majorana state from the rest of the vortex core bound states, where $R$ is the radius of the hole (for a pinned vortex, $R$ is replaced with the superconducting coherence length). This mini-gap usually ranges in the sub-meV. In addition, one must resolve the gap that separates the unpaired Majorana bound state from the rest of the spectrum in the Majorana chain along the vortex line. This gap goes as $\sim |\Delta|-|\im t|$ and is shown in the grayscale of Fig.~\ref{fig:phased}. In our estimate, this gap would also range in the sub-meV. However, the weight of the states above this gap is greatly diminished since they are extended along the hole and therefore have a smaller overlap with the tunneling tip. Overall, sub-meV energy resolution and similarly low temperatures would be typically needed for such an experiment. 

These authors have recently proposed~\cite{GroSerVis10a} an interference experiment based on the Aharonov-Casher effect to detect the non-Abelian statistics associated with unpaired Majorana fermions. The present work shows that a setup similar to that shown in Fig.~\ref{fig:exp} with unpaired Majorana fermions bound to the top and bottom layers is more suitable for such an experiment. We note that the ability to grow high-quality thin film samples of Sr$_2$RuO$_4$ in a layer-by-layer fashion has been recently demonstrated~\cite{KroUchTak10a}. As our discussion shows, such a degree of control will be very useful in the clear identification of Majorana fermions and their potential applications in layered superconducting material.

%--- Conclusion
\section{Conclusion}

We studied the problem of Majorana modes on a vortex line threading a layered two-component topological superconductor. We mapped the problem to that of a general quantum wire (or a Majorana chain with two Majorana states per site) with nearest neighbor pairing term and a finite supercurrent. We established both analytically and numerically that the topological phase of the quantum wire, which supports unpaired Majorana bound states at the boundaries of the chain, survives in the presence of a supercurrent when the latter is below the critical current of the wire. This generalizes the condition derived by Kitaev for unpaired Majorana modes~\cite{Kit00a} (that the quantum wire be partially filled). 

For a spin-triplet chiral p-wave superconductor we calculated the Majorana tunneling amplitudes due to spin-rotation breaking interactions and derived the conditions for which unpaired Majorana modes may be stabilized in a vortex line at the top and bottom layers. Our study suggests that unpaired Majorana bound states could exist even when Majorana tunneling in and between layers is nonzero if a supercurrent flows between the layers and for certain spin-orbit couplings. These vortices satisfy non-Abelian braiding statistics with half-quantum vortices when there is an odd number of layers. More generally, our results suggest that the supercurrent could be useful as an additional control knob for realizing unpaired Majorana fermions in one-dimensional systems.

%--- Acknowledgment
\section*{Acknowledgment}
This work is supported by the Institute for Condensed Matter Theory at University of Illinois, Urbana-Champaign. The authors acknowledge useful discussions with S. Vishveshwara and E.~Y. Andrei.

\appendix

%---- Majorana modes in a single layer
\section{Majorana fermions in a chiral p-wave superconductor}
\label{app:Mmodes}

\subsection{Majorana edge modes}
For a spin-polarized $p_x+  i  p_y$ condensate,
\begin{eqnarray}\label{eq:spinless}
	\Delta= i  v_\Delta  e^{ i \Phi/2}(\partial_x-  i  \partial_y)  e^{ i \Phi/2}.
\end{eqnarray}
This condensate has a single Majorana mode in each layer at the boundary of the system normal to the xy-plane $\vex r= (r,\theta)$. For a disk of radius $R$, threaded by $n$ vortices, i.e. when $\Phi=n\theta$, the energy of these modes is $E=(v_\Delta/R)l$ (see Appendix~\ref{app:Mmodes} for details) where the angular momentum $l\in\mathbb{Z}$ for odd $n$ and $l\in\mathbb{Z}+\frac12$ for even $n$. Thus, for a single vortex there is a single zero energy edge state for any $R$. %Fig.~\ref{fig:exp} depicts the modes in an annulus geometry.

We take
%%%%%%
\begin{eqnarray}
	h=-\frac{1}{2m}\boldsymbol{\nabla}^2+\frac{1}{2 m_z} p_z^2+V_{\rm c}(r)+V_{\rm p}(z)-\mu_{\rm c},
\end{eqnarray}
%%%%%%
with the confining potential $V_{\rm c}(r)$ chosen for a disk of radius $R$ to be $V_{\rm c}(r\ll R)=0$ and $V_{\rm c}(r\gg R)=V_0$ where $V_0$ is large and positive. The periodic potential $V_{\rm p}(z)$  creates the layered structure: the layers are potential minima at $z=\ell a$ where $\ell$ is an integer. We first concentrate on a single layer, i.e. we solve $\left[p_z^2/2m+V_{\rm p}(z)\right]\psi_{z,\ell}(z)=E_z\psi_{z,\ell}(z)$ around each $z=\ell a$ and project to the lowest energy state. Next, we define the electrochemical potential $\varepsilon(r)=\mu_{\rm c}-V_{\rm c}(r)-E_z$, and choose $\varepsilon(R)=0$ so that $\varepsilon$ will be negative on one side of the edge (the insulator at $r>R$) and positive on the other (the topological superconductor at $r<R$). Let us specialize to the spin-polarized chiral p-wave condensate in~(\ref{eq:spinless}). Linearizing the Hamiltonian near $r=R$, taking the phase of the order parameter to be $\Phi=n \theta$ where $n$ is the number of vortices through the superconductor, and choosing the angular dependence of the spinor to be $\exp\left[ i \sigma_z (n-1)\theta/2\right]$, we can factorize the BdG Hamiltonian as
%%%%%
\begin{eqnarray}
	\label{eq:linearized-HH}
	H=-\tau_z \left[\varepsilon(r)+v_\Delta \tau_y\left(\partial_r+\frac{1}{2r}\right)\right]+ i  v_\Delta \tau_y \frac{1}{R}\partial_\theta.\nonumber \\
	~
\end{eqnarray}
%%%%%
The radial part contains a zero mode, $\left[\varepsilon(r) - v_\Delta\left(\partial_r+\frac{1}{2r}\right)\right]f_{\rm e}=0$,
\begin{eqnarray}
	 f_{\rm e} =\frac{1}{\sqrt{r}}\exp\left[+\frac{1}{v_\Delta}\int_R^r\varepsilon(\rho) d \rho\right].
\end{eqnarray}
We assume that the potential is sharp enough so that higher energy states are not important. We can now write the Nambu spinor projected into this zero energy state
\begin{eqnarray}
	\Psi_{0}({\bf r})= e^{ i \tau_z \phi/2} e^{ i \tau_z(n-1)\theta/2} e^{ i \pi/4} f_{\rm e}(r) \psi(\theta) \left(\begin{array}{c}1 \\ -  i  \end{array}\right),\nonumber\\
	~
\end{eqnarray}
where $\phi=\Phi-n\theta$ is a constant. The effective Hamiltonian for the Hermitian field $\psi(\theta)$ is $H_{0}= -  i (v_\Delta/R) \partial_\theta$. We can expand $\psi(\theta)=\sum_l  e^{ i  l \theta}\psi_l$, where $\psi_l$ satisfies $\psi_l^\dag=\psi_{-l}$. The energy levels are $E_l=-(v_\Delta/R) l$. We note that the spinors are single-valued only if $l\pm(n-1)/2\in \mathbb{Z}$. Consequently, if $n$ is odd, then $l\in \mathbb{Z}$, and if $n$ is even, $l\in \mathbb{Z}+1/2$.

As is apparent from equation (\ref{eq:linearized-HH}), the eigenvalue of $\sigma_y$ is chosen by the asymptotic behavior of the electro-chemical potential $\varepsilon$, locking the direction of propagation of the edge to the spatial profile of $\sgn(\varepsilon(r))$.

\subsection{Majorana bound states}

For the HQV in the spin-triplet chiral p-wave condensate in~(\ref{eq:triplet}), we take
$
\vex{d}(r,\theta)=\left(\cos\frac{\theta}{2},-\sin\frac{\theta}{2},0\right),
$
and $\Phi=\phi+\theta/2$ with a constant $\phi$.
Since
%%%%%
\begin{equation}
 e^{ i \theta/2} (\boldsymbol{\sigma}\cdot\vex{d})  i   \sigma_{y}=\left(
\begin{array}{cc}
-  e^{ i \theta} & 0\\
0 & 1 \end{array}\right),
\end{equation}
%%%%%
only the spin-up component sees a vortex winding in the phase. The BdG equation has a zero mode
%%%%%
\begin{eqnarray}
\chi_{\uparrow} &=& e^{ i  \sigma_z \phi/2} e^{ i \pi/4} f \left(
\begin{array}{c}
v_\uparrow \\
- i  v_\uparrow
\end{array}
\right),\\
f(r) &=& \mathcal{N} e^{-mv_\Delta r} \left\{ \begin{array}{c}
J_0(\kappa r) \quad \varepsilon >\frac12mv_\Delta^2, \\ 
I_0(\kappa r) \quad \varepsilon <\frac12mv_\Delta^2,
\end{array}\right.
\end{eqnarray}
%%%%%
where $v_\uparrow = (1,0)^T$, $\kappa\equiv \sqrt{2m|\varepsilon-\frac12mv_\Delta^2|}$ and $\mathcal{N}$ is a normalization factor. Here, we also assumed a vanishing small vortex core so that $v_\Delta$ is constant everywhere but at the origin, as well as a constant $\varepsilon$ away from the edges.

The $\vex d$ vector in the presence of a FQV does not contain any winding and we simply take it to be a constant. The Hamiltonian is invariant under a SU(2) spin rotation $\Psi\to U\Psi'$, with $U=\left(\begin{array}{cc} S & 0 \\ 0 & S^* \end{array}\right)$, $S\in$ SU(2), and $\vex d\to R_S^{-1} \vex d'$, where $R_S\in$ SO(3) is the rotation corresponding to $S$. With this symmetry we can rotate $\vex d \mapsto \hat{\vex y}$, for which the Hamiltonian decouples into spin components with a vortex phase winding $\Phi=\phi+\theta$ in both spin components. There are two zero modes
%%%%%%
\begin{equation}\label{eq:spinors}
\chi'_{\sigma}= e^{ i  \sigma_z \phi/2} e^{ i \pi/4} f \left(
\begin{array}{c}
v_\sigma \\
- i  v_\sigma
\end{array}
\right),
\end{equation}
%%%%%
with $\sigma=\;\uparrow, \downarrow$ and $v_\downarrow=(0,1)^T$. Rotating back to the original basis, the full solutions are found as $\chi_\sigma = U \chi'_\sigma$.

\section{Majorana tunneling}
\label{app:tun}

\subsection{Intra-layer Majorana tunneling}

The BdG Hamiltonian receives a contribution $H_{\rm Z}$ from Zeeman interaction, which after the rotation $R_S: \vex d\mapsto \hat{\vex y}$ reads
%%%%%
\begin{equation}
	U^\dagger H_{\rm Z} U=\left(\begin{array}{cc}\mu_{\rm B}\,S^\dagger\boldsymbol{\sigma} \cdot {\vex B} S & 0 \\ 0 & -\mu_{\rm B}(S^\dagger \boldsymbol{\sigma} \cdot {\vex B} S)^T \end{array}\right).
\end{equation}
%%%%%
So, using the identity
%%%%%
\begin{equation}
S^\dagger\boldsymbol{\sigma} \cdot {\vex B} S = \boldsymbol{\sigma} \cdot {R_S \vex B}
\end{equation}
%%%%%
we have
%%%%%
\begin{equation}
	\alpha=- i \langle \chi'_\uparrow|U^\dagger H_{\rm Z} U|\chi'_\downarrow\rangle=-2\mu_{\rm B} \vex B \cdot \vex d.
\end{equation}
%%%%%

The spin-orbit term when $\boldsymbol\lambda  || \hat{\vex z}$ was considered in Ref.~\cite{LuYip08a} and found not to lift the degeneracy of the two Majorana modes. We show that, to the first order in perturbation theory, this is actually true for general $\boldsymbol\lambda $. We have
%%%%%
\begin{equation}
H_{\rm so} = \left(
\begin{array}{cc}
\boldsymbol{\lambda }\cdot(\boldsymbol{\sigma}\times\vex p) & 0 \\
 0 & -\boldsymbol{\lambda }\cdot(\boldsymbol{\sigma}\times\vex p)^T 
\end{array}
\right).
\end{equation}
%%%%%
Therefore, the overlap $\langle \chi_\uparrow|H_{\rm so}|\chi_\downarrow\rangle$ will contain terms involving components of the integral $\int (f\vex p f)  d \vex r \propto \int (\boldsymbol{\nabla} |f|^2) d \vex r = 0$ since $f$ vanishes at infinity. We conclude
%%%%%
\begin{equation}\label{eq:intraSO}
\langle \chi_\uparrow|H_{\rm so}|\chi_\downarrow\rangle = 0.
\end{equation}
%%%%%

When $\vex B\cdot\vex d=0$, the Zeeman tunneling vanishes, which is true to all orders of perturbation theory. The chemical potential is shifted for each spin component to $\mu_\pm = \mu_{\rm c} \pm \mu_{\rm B} B$. Since the two spin blocks are decoupled, we still find two zero modes but with new functions $f_\pm$. If we then turn on the spin-orbit interaction, to the first order of perturbation theory, $\int f_+ (\boldsymbol\lambda \times\boldsymbol{\nabla})_x f_-  d \vex r = 0$ since $\int f_+\partial_z f_- d \vex r = 0$ and $\int f_+\partial_y f_- d \vex r = 0$ by cylindrical symmetry of $f_\pm$. So~(\ref{eq:intraSO}) still holds.

\begin{figure}[tb]
\begin{center}
\includegraphics[width=2.25in]{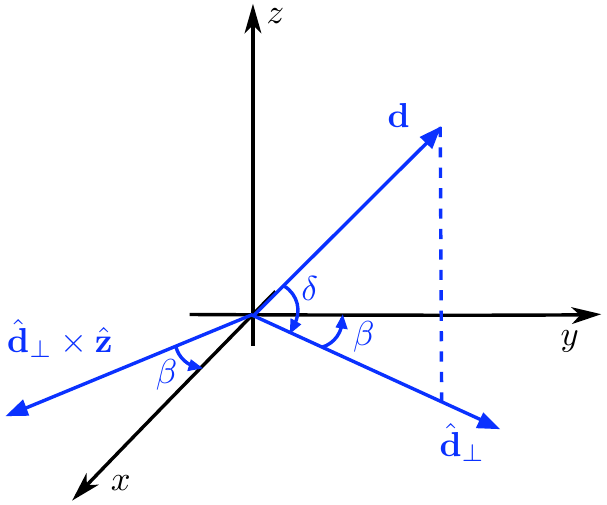}
\end{center}
\caption{(color online) The geometry of the $\vex d$ vector.}
\label{fig:ddperp}
\end{figure}

\subsection{Interlayer Majorana tunneling}
Let us consider the overlap between the Majorana modes of a FQV in two adjacent layers indexed as 1 and 2. It is given by $t = - i  \left<\chi_2|H_{\rm o}|\chi_1\right>$, where $H_{\rm o} = \varepsilon_0\left(\begin{array}{cc} I & 0 \\ 0 & - I \end{array} \right)$ with $\varepsilon_0$ an energy scale for the tunneling process. So, by defining
%%%%%%
\begin{equation}
t_0 = 2 \int f_2(\vex r) f_1(\vex r) d \vex r  d  z,
\end{equation}
%%%%%%
we find $(t_{\rm o})_{\sigma\overline{\sigma}} = 0$ and
%%%%%%
\begin{equation}\label{eq:to}
(t_{\rm o})_{\sigma\sigma} = t_0 \varepsilon_0 \sin\varphi,
\end{equation}
%%%%%%
where $\varphi = (\Phi_1 - \Phi_2)/2$ is half of the superconducting phase difference between the layers. So, in a uniform condensate $t_{\rm o}=0$, even though Cooper pairs move freely in all three direction! 

The Zeeman interaction yields the tunneling amplitudes
%%%%%%
\begin{eqnarray}
\left(t_{\rm Z}\right)_{\sigma\sigma'} &=& - i  \left<\chi'_{2\sigma}| U^\dagger H_{\rm Z} U | \chi'_{1\sigma'} \right> \nonumber \\ 
&=&  -\frac i 2 t_0 \mu_{\rm B} \vex B \cdot v_\sigma^T \left[  e^{ i \varphi} S^\dagger \boldsymbol{\sigma} S
-  e^{- i \varphi} (S^\dagger \boldsymbol{\sigma} S)^T \right] v_{\sigma'} \nonumber \\
&=& t_0 \mu_{\rm B} \vex B \cdot {\rm Im}\left(  e^{ i \varphi} v_\sigma^T S^\dagger \boldsymbol{\sigma} S v_{\sigma'}\right).
\end{eqnarray}
%%%%%%

We shall use the following parameterization $\vex d = (\cos\delta\sin\beta, \cos\delta \cos\beta, \sin\delta)$ visualized in Fig.~\ref{fig:ddperp} and define $\hat{\vex d}_\perp = (\sin\beta, \cos\beta,0)$ as the unit vector along the projection of $\vex d$ onto the xy-plane. Then $S=\exp( i  \beta \sigma_z/2)\exp(- i \delta \sigma_x/2)$.
%%%%%%%
%\begin{eqnarray*}
%v_\sigma^T \sigma_x v_{\sigma'} = v_\sigma^T \sigma_x^T v_{\sigma'} = \delta_{\overline{\sigma}\sigma'},\\
%v_\sigma^T \sigma_y v_{\sigma'} = - v_\sigma^T \sigma_y^T v_{\sigma'} = - i  (-1)^{\sigma} \delta_{\overline{\sigma}\sigma'},\\
%v_\sigma^T \sigma_z v_{\sigma'} = v_\sigma^T \sigma_z^T v_{\sigma'} =  (-1)^{\sigma} \delta_{\sigma\sigma'},
%\end{eqnarray*}
%%%%%%%
It is straightforward to find
\begin{widetext}
%%%%%%
\begin{eqnarray}
(t_{\rm Z})_{\sigma\sigma} &=& (-1)^\sigma t_0 \mu_{\rm B} (- B_x \sin\delta\sin\beta - B_y \sin\delta\cos\beta + B_z \cos\delta) \sin \varphi \\ %= (-1)^\sigma t_0 \mu_{\rm B} \vex B \cdot [ (\hat{\vex d}_\perp \times \hat{\vex z} \times \vex d] \sin\varphi \\
(t_{\rm Z})_{\sigma\overline{\sigma}} &=& -(-1)^\sigma t_0 \mu_{\rm B} (B_x \cos\delta\sin\beta + B_y \cos\delta \cos\beta + B_z \sin\delta) \cos\varphi + t_0 \mu_{\rm B} (B_x \cos\beta - B_y \sin\beta) \sin\varphi,
\end{eqnarray}
%%%%%%

Substituting back the components of $\vex d$ and $\hat{\vex d}_\perp$ we have then
%%%%%%
\begin{eqnarray}\label{eq:tZa}
\left(t_{\rm Z}\right)_{\sigma\sigma} &=&  (-1)^{\sigma} t_0 \mu_{\rm B} \left[\hat{\vex d}_\perp \times (\vex B \times \vex d)\right]_z \sin\varphi,\\
\label{eq:tZb}
\left(t_{\rm Z}\right)_{\sigma\overline{\sigma}} &=& t_0 \mu_{\rm B}
 \left[ (\vex B \times \hat{\vex d}_\perp)_z \sin\varphi - (-1)^{\sigma} \vex B\cdot \vex d \cos\varphi  \right].
\end{eqnarray}
%%%%%%

The contribution from the spin-orbit interaction can be similarly written as
%%%%%%
\begin{eqnarray}\label{eq:interSO}
\left(t_{\rm so}\right)_{\sigma\sigma'} &=& - i  \left<\chi'_{2\sigma}| U^\dagger H_{\rm so} U | \chi'_{1\sigma'} \right> \nonumber \\ 
&=&  v_\sigma^T \left[  e^{ i \varphi} S^\dagger \boldsymbol{\sigma} S
+  e^{- i \varphi} (S^\dagger \boldsymbol{\sigma} S)^T \right] v_{\sigma'} \cdot \int f_2 (\boldsymbol\lambda  \times \boldsymbol{\nabla}) f_1   d \vex r  d  z. \nonumber \\
&=& 2 {\rm Re} \left(  e^{ i \varphi} v_\sigma^T S^\dagger \boldsymbol{\sigma} S v_{\sigma'} \right) \cdot \int f_2 (\boldsymbol\lambda  \times \boldsymbol{\nabla}) f_1   d \vex r  d  z.
\end{eqnarray}
%%%%%%
%So,
%\begin{eqnarray}
%\left(t_{\rm so}\right)_{\sigma\sigma'} &=& 
%2 \delta_{\overline{\sigma}\sigma'} \cos\left(\frac{\Delta\Phi}2\right) \int  f_2 (\boldsymbol\lambda  \times \boldsymbol{\nabla})_x f_1 d\vex r dz \nonumber \\
%&+& (-1)^{\sigma} 2 \delta_{\overline{\sigma}\sigma'} \sin\left(\frac{\Delta\Phi}2\right) \int  f_2 (\boldsymbol\lambda  \times \boldsymbol{\nabla})_y f_1 d\vex r dz \nonumber \\
%&+& (-1)^{\sigma} 2 \delta_{\sigma\sigma'} \cos\left(\frac{\Delta\Phi}2\right) \int f_2 (\boldsymbol\lambda  \times \boldsymbol{\nabla})_z f_1  d\vex r dz .
%\end{eqnarray}
%%%%%%%
Assuming cylindrical symmetry (i.e. that the vortex does not wiggle), we have $\int f_2(\partial_x,\partial_y) f_1 d \vex r = 0$. Then~(\ref{eq:interSO}) simplifies to
%%%%%%
\begin{eqnarray}
(t_{\rm so})_{\sigma\sigma} &=& (-1)^\sigma t'_0 (\lambda_{{\rm so},x} \sin\delta\cos\beta - \lambda_{{\rm so},y} \sin\delta\sin\beta  ) \sin\varphi \\
(t_{\rm so})_{\sigma\overline{\sigma}} &=& t'_0 (\lambda_{{\rm so},x} \sin\beta + \lambda_{{\rm so},y} \cos\beta  ) \cos\varphi + (-1)^\sigma t'_0 (-\lambda_{{\rm so},x} \cos\delta\cos\beta + \lambda_{{\rm so},y} \cos\delta\sin\beta) \sin\varphi,
\end{eqnarray}
%%%%%%
with
%%%%%%
\begin{eqnarray}
t'_0 = 2 \int  f_2 \partial_z f_1  d \vex r  d  z.
\end{eqnarray}
%%%%%%
Then
%%%%%%
\begin{eqnarray}\label{eq:tsoa}
\left(t_{\rm so}\right)_{\sigma\sigma} &=& (-1)^\sigma t'_0  \hat{\vex d}_\perp \cdot(\vex d \times \boldsymbol\lambda ) \cos\varphi,  \\
\label{eq:tsob}
\left(t_{\rm so}\right)_{\sigma\overline{\sigma}} &=&
t'_0 \left[ \boldsymbol\lambda \cdot\hat{\vex d}_\perp \cos\varphi + 
(-1)^{\sigma} (\vex d \times \boldsymbol\lambda )_z \sin\varphi \right].
\end{eqnarray}
%%%%%%
\end{widetext}

%-----------------

\end{document}